\documentclass[aps,pre,onecolumn,longbibliography]{revtex4-2}
\usepackage{graphicx}
\usepackage{xcolor}
\usepackage{amsmath}
\usepackage[colorlinks=true,hyperfootnotes=true,breaklinks=true]{hyperref}

\DeclareMathOperator{\Prob}{\mathrm{Prob}}

\begin{document}

\title{ 
Heider balance in square versus triangular lattices under an external
field}

\author{Zdzis{\l}aw Burda}
\email{zdzislaw.burda@agh.edu.pl}
\thanks{ORCID~\href{https://orcid.org/0000-0002-9656-9570}{0000-0002-9656-9570}}

\author{Maciej Wo{\l}oszyn}
\email{woloszyn@agh.edu.pl}
\thanks{ORCID~\href{https://orcid.org/0000-0001-9896-1018}{0000-0001-9896-1018}}

\author{Krzysztof Malarz}
\email{malarz@agh.edu.pl}
\thanks{ORCID~\href{https://orcid.org/0000-0001-9980-0363}{0000-0001-9980-0363}}

\author{Krzysztof Ku{\l}akowski}
\email{kulakowski@fis.agh.edu.pl}
\thanks{ORCID~\href{https://orcid.org/0000-0003-1168-7883}{0000-0003-1168-7883}}

\affiliation{\href{https://ror.org/00bas1c41}{AGH University},
Faculty of Physics and Applied Computer Science,
al. Mickiewicza 30, 30-059 Krak\'ow, Poland}

\date{September 20th, 2026}

\begin{abstract}
We discuss the Heider model in the presence of an external social field. This field breaks the possible symmetry between the probabilities of hostile and friendly relationships. We consider the system in the presence of fluctuations generated by thermal noise and present the results of a comparative study of two-dimensional triangular and square networks with periodic boundary conditions. The results were obtained using three different methods: exact calculations for small systems ($3 \times 3$ square lattice and $2 \times 2$ triangular lattice), Monte Carlo simulations of medium-sized systems ($16 \times 16$ lattices), and exact calculations in the thermodynamic limit (corresponding to infinite size) of certain limiting cases for which analytical solutions are possible. In particular, we exploit the recently discovered equivalence between structurally balanced systems and the Ising model to derive an exact form of the edge magnetization susceptibility for systems in Heider equilibrium.
\end{abstract}

\maketitle

\section{Introduction}

The Heider balance, or structural balance \cite{Heider,Cartwright_1956}, is a vast field of research in various disciplines \cite{Crandall_2011,2010.10036,Kulakowski_2025,Chiang_2020}.
Initially formulated for complete graphs \cite{Antal_2006,PhysRevE.105.054312,1911.13048}, models of dynamics leading to balance have been applied to networks of various topologies, such as triangular lattices \cite{2007.02128} (also diluted and enhanced \cite{2106.03054}), classical random graphs \cite{2206.14226} and Archimedean lattices of different kinds \cite{2407.02603}. 
In all these attempts, triads of mutually connected nodes remain the basic units to which the idea of balance can be applied. 
Briefly, a triad is balanced if the number of its negative links is even \cite{Cartwright_1956}.
The special case of absence of these negative links in the whole system is termed as `paradise'. 

The sociological basis of the Heider balance is to evade cognitive dissonance, which appears when one is not able to determine who is a friend and who is an enemy. 
The method is often formulated as a verification if all triads are balanced, i.e. if the number of negative links in each triad is even. 
A more general way is to check if this number is even in each closed path. This is equivalent to an attempt to identify two communities that are internally friendly but mutually hostile. 
Actually, the existence of such a partition is the criterion of balance, either with or without field \cite{Cartwright_1956,2511.17247}. The latter criterion does not depend on the network topology, so it works also for the square lattice.

We note that a sociological example of a square lattice is the Schelling model \cite{Schelling_1971} 
 of residential segregation in American neighborhoods, where lattice cells represent housing units.
It indicates that the topology of a square lattice is commonly accepted as an example of a social network.

Our aim here is to explore the Heider balance in a square lattice and in the presence of an external field. The concept of a social field is useful in social psychology \cite{Helbing_1994,Helbing_1995,Helbing_2010}, as it represents external incentives to improve or deteriorate interpersonal relationships. 
The concept evolved: as formulated by Kurt Lewin, the social field is a space of socially
determined individual and social propensities to some actions \cite[p. 197]{Lewin_1951}. The conditions of this or that action include public 
opinion, social norms and expectations as well as a psychological state of an individual. In a more mathematized formulation 
\cite{Helbing_1994}, a change of state $x$ of an individual can be found from a gradient of some social field $V(x)$. 
Accordingly, a tension to change vanishes when $V(x)$ assumes a minimum. 
In a recent literature (References \onlinecite{Oloomi_2023,2512.02644} and the present text) 
the social field to be minimized is equivalent to a magnetic field $h$ acting on spins attached to links of the social network. 
Then, a positive (negative) value of $h$ marks a preference of friendly (hostile) relations between connected agents, and for 
large absolute value of $h$ this preference is stronger. Although these mathematical formulations are different, a common crux can be seen 
in their social interpretations. The social field represents the state of social environment, which influences the probability 
of a given behavior. In our case, all factors should be included which contribute to the signs and values of the interpersonal relations.
To be concrete, we can refer to an example. In different countries, an agreement was measured with the statement: `most people can be trusted'
\cite{link1}. The differences are huge: from 74 percent of support in Denmark, down to 4 percent in Peru \cite{link2}. These numbers---a conformation to the social norm of mutual trust---can be freely interpreted as an important component of the social 
field.

The novelty here is the test of the effect of an external field on the standard triad-filled topology and the triad-free topology, where the elementary cycles are quadruples. 
The difference is essential. For triad-filled topology, a negative field $h$ competes with the Heider mechanism, while for the square lattice, it does not. Namely, in the latter case, the paradise state with all positive edges 
has a counterpart: the `anti-paradise' (or `hell') balanced state where all edges are negative.

A convenient way to implement Heider theory is to define the parity of a path (in a relationship network) as the product of the signs of the links that belong to the path \cite{Cartwright_1956}. In a balanced network, the parity of a path between friends is positive, while that between enemies is negative, regardless of the complexity of that path. Furthermore, each closed path has a positive parity. 
Moreover, the set of friends of a node consists of all nodes accessible from this node via positive-parity paths. In addition, if each elementary loop is balanced, then the whole graph \cite{Cartwright_1956} is also balanced.

Quadruples in the context of Heider balance have been used in Reference~\onlinecite{Hao_2024} for the construction of link configurations with appropriate randomization. There, the randomization rules preserve both the network topology and the signed degrees. Larger structures have not been considered there. In addition, quadruples appear in References~\onlinecite{Kargaran_2020,Hakimi_2022} as a consequence of the postulated interaction of neighboring triads, with one link shared. We are unaware of any study of the structural balance for networks entirely devoid of triads.

In the case of triads, the structural balance condition is satisfied by one of four cases: a friend of my friend is my friend; or a friend of my enemy is my enemy; or an enemy of my friend is my enemy; or an enemy of my enemy is my friend. Applied to the quadruple $abcd$, which forms an elementary loop on a square lattice in which $a$ has two immediate neighbors, $b$ and $d$; and one $c$ that is not a neighbor, the corresponding balance condition, viewed from the perspective of node $a$, is satisfied by one of six cases:
a friend $c$ of my friend $b$ is a friend of my friend $d$; 
or a friend $c$ of my friend $b$ is an enemy of my enemy $d$; 
or a friend $c$ of my enemy $b$ is a friend of my enemy $d$; 
or a friend $c$ of my enemy $b$ is an enemy of my friend $d$; 
or an enemy $c$ of my friend $b$ is an enemy of my friend $d$; 
or an enemy $c$ of my enemy $b$ is an enemy of my enemy $d$.

\section{The model}

When modeling social phenomena, statistical mechanics provides a set of invaluable concepts \cite{Sen_2014}. 
We will apply them to understand the model, keeping in mind the sociological context. The basic sociological assumption is that people tend to minimize their cognitive dissonance \cite{Festinger_1957,Festinger_1962}. This process is encoded as the minimization of energy. The process is stochastic in nature, and this will be captured in the model by the presence of thermal noise. The nature of social interactions will be modeled by an external field $h$ that increases the probability of friendly ($h>0$) or hostile relationships ($h<0$). 
The system's tendency to achieve structural balance will be controlled by a parameter $\varepsilon$: the larger $\varepsilon$, the easier the balance is achieved. In the language of statistical physics, this comes down to the following expression for the energy of the system
\begin{equation}
\label{U}
   U = -\varepsilon P - h M, 
\end{equation}
where
\begin{equation}
\label{P}
   P = \frac{1}{n_p} \sum_{\{abcd\}}  s_{ab} s_{bc} s_{cd} s_{da}    
\end{equation}
and 
\begin{equation}
\label{M}
M = \frac{1}{n_e} \sum_{\{ab\}} s_{ab}. 
\end{equation}
The sign $s_{ab}$ is positive, $s_{ab}=1$, indicating a friendly relationship between $a$ and $b$, and negative, $s_{ab}=-1$, indicating a hostile relationship between $a$ and $b$. 
The sum \eqref{P} runs over elementary loops, which are quadruples $\{abcd\}$ for a square lattice or triads $\{abc\}$ for a triangular lattice,
in which case 
\[P = \frac{1}{n_p} \sum_{\{abc\}}  s_{ab} s_{bc} s_{ca}.\]
The sum \eqref{M} runs over the edges $\{ab\}$. The parameters $n_p$ and $n_e$ are the numbers of elementary loops (plaquettes) per node and the number of edges per node, respectively;  i.e., $n_p=N_p/N$, $n_e=N_e/N$, where $N_p,N_e,N$ are the numbers of elementary loops (plaquettes), edges, and nodes in the network forming the lattice. 
The factors $n_p$ and $n_e$ used in the definitions of $P$ and $M$ transform averages of the signs of elementary loops and edges into the densities per node $P/N= \sum_p (ssss)_p/N_p$ and $M/N = \sum_{e} s_e /N_e$. Thus, in particular, dividing both sides of Eq.~\eqref{U} by $N$ yields an equation that relates densities per node, having a natural physical interpretation.

For the given temperature $T$, the system is described by the partition function 
\begin{equation}\label{Z}
Z(\beta,\varepsilon,h,N)=\sum_{\{s_{ab}\}} \exp(-\beta U) = 
\sum_{\{s_{ab}\}} \exp(\beta \varepsilon  P + \beta h M),
\end{equation}
where $\beta$ is the inverse temperature, $\beta=1/T$, and the sum is over all states represented by a collection of signs $s_{ab}=\pm 1$ for all edges. 
In what follows, we will consider 2D toroidal lattices $L \times L$, having $N=L^2$ nodes. 
For a square toroidal lattice $N_p = N$ and $N_e=2N$, hence $n_p=1$ and $n_e=2$. 
For a triangular toroidal lattice $N_p = 3N$ and $N_e=2N$, hence $n_p=2$ and $n_e=3$.
For the given size $N$, the partition function 
\begin{equation} \label{Z'}
Z'(\varepsilon',h',N)= \sum_{\{s_{ab}\}} 
\exp\left(\varepsilon'  P + h' M\right)
\end{equation}
effectively depends on two independent parameters $\varepsilon'=\beta \varepsilon=\varepsilon/T$ and $h'=\beta h = h/T$.
In the limit $\varepsilon' \rightarrow \infty$, the system reaches a structural balance, in which the parity of each elementary loop is equal to one. The parity $\pi(\gamma)$ of a path $\gamma=\{a_1,a_2,\ldots, a_{n+1}\}$ that consists of $n$ edges $a_i a_{i+1}$ is defined as
\begin{equation} \label{parity}
 \pi(\gamma) = \prod_{i=1}^{n} s_{a_i a_{i}}.
\end{equation}
In a structurally balanced system, the parity of each elementary loop $\gamma_{e} = \{a,b,c,d,a\}$ in a square lattice and $\gamma_{e}=\{a,b,c,a\}$ in a triangular lattice is $\pi(\gamma_e)=1$. 
An immediate consequence is that for any loop $\gamma_{l}$ in the structurally balanced system, the parity is positive $\pi(\gamma_{l})=1$. 
A further consequence is that all open paths $\gamma_{ab}$ having the same endpoints $a$ and $b$ have the same parity in a structurally balanced system: $\pi(\gamma_{ab})=\pi(\gamma'_{ab})$. This means that, in this case, the parity of a path $\gamma_{ab}$ depends only on the endpoints. As a consequence, for a given $a$, one can easily determine all $b$'s for which $\pi(\gamma_{ab})=1$, and all $b$'s for which $\pi(\gamma_{ab})=-1$, which gives a clear division of the system into friends and foes of $a$. In a structurally balanced system, there is still a dependence on the social field $h$, which breaks the symmetry between the probabilities of friendly and hostile interactions. The question is what the effect of this field will be on the system.

\section{Observables}
The main physical quantity is free energy
\begin{equation} 
\label{F}
 F(\beta,\varepsilon,h,N) = -{\beta}^{-1} \ln Z(\beta,\varepsilon,h,N),
\end{equation}
which can be calculated from the partition function $Z$ \eqref{Z}.
Taking derivatives of $F$ with respect to $\varepsilon$ and $h$, one can calculate the statistical means of $P$ and $M$, and by taking higher-order derivatives, one can calculate the correlations. More precisely, free energy is a generating function for cumulants. The first order cumulants give the means
\begin{subequations}
\begin{equation}
     \langle P \rangle = -\partial_\varepsilon  F ,
     \end{equation}
\begin{equation}
    \langle M \rangle = -\partial_h  F.
\end{equation}
\end{subequations}
The second order cumulants give the covariance matrix:
\begin{subequations}
\begin{equation}
    \langle P^2 \rangle - \langle P \rangle^2  = - \beta^{-1} \partial^2_{\varepsilon\varepsilon}  F, 
\end{equation}
\begin{equation}
    \langle M^2 \rangle - \langle M \rangle^2  = - \beta^{-1} \partial^2_{hh}  F,  
\end{equation}
and 
\begin{equation}
\langle P M \rangle - \langle P \rangle \langle M \rangle =
    - \beta^{-1} \partial^2_{\varepsilon h}  F.
\end{equation}
\end{subequations}
In the thermodynamic limit ($N\to\infty$) the free energy per node approaches an $N$-independent density
\begin{equation} \label{fe_density}
    f(\beta,\varepsilon,h) = \lim_{N\rightarrow \infty} \frac{F(\beta,\varepsilon,h,N)}{N}.
\end{equation}
Since all the cumulants are derivatives of free energy, which asymptotically grows linearly with $N$, they should also asymptotically grow linearly with $N$ when $N$ approaches infinity. The exceptions are phase transition points that correspond to specific parameter values, known as critical values, where the higher-order derivatives of the free energy are infinite. An example is a second-order phase transition in which the second derivatives are discontinuous or divergent. 
In this case, the second- and/or higher-order cumulants grow faster than linearly with $N$, signaling the presence of long-range correlations in the system. The second order phase transition takes place at the parameters $\beta,\varepsilon,h$, where any second order derivative of the density \eqref{fe_density} is divergent or discontinuous. If the analytic expression for the free energy density is known, one can determine the critical values of the parameters analytically. This is usually not the case. To find the phase transition and determine its order,
one must usually rely on numerical calculations. The standard way to check whether critical points exist is to perform a finite-size analysis. 
This is done by checking whether the susceptibilities (normalized second-order cumulants)
grow to infinity with $N$ at some values $\beta$, $\varepsilon$, and $h$. The first step is to compute the mean values of elementary
loop signs 
\begin{subequations}
\label{eq:pm}
\begin{equation} 
\label{eq:p}
    p = \frac{\langle P \rangle}{N} = -\partial_\varepsilon f
\end{equation}
and edge magnetization
\begin{equation} \label{eq:m}
    m = \frac{\langle M \rangle}{N} = - \partial_h f.
\end{equation} 
\end{subequations}
Then, one computes the second order cumulants, which correspond to susceptibilities:
\begin{subequations}
\label{eq:chis}
\begin{equation} {\label{eq:chip}}
    \chi_p =  \partial_\varepsilon p = - \partial^2_{\varepsilon\varepsilon}  f = \frac{\beta}{N} \left(\langle P^2 \rangle - \langle P \rangle^2\right);
\end{equation}
\begin{equation} {\label{eq:chim}}
    \chi_m = \partial_h m = - \partial^2_{hh}  f = \frac{\beta}{N} \left(\langle M^2 \rangle - \langle M \rangle^2\right) ; 
\end{equation}
and 
\begin{equation}
    \chi_{pm} = -\partial^2_{\varepsilon h}  f = \frac{\beta}{N} \left(\langle PM \rangle - \langle P \rangle \langle M \rangle\right) .
\end{equation}
\end{subequations}
More precisely, one can numerically calculate the quantities on the right-hand side of the above equations and check how they behave
when $N$ increases. If any susceptibility grows linearly with $N$ at certain values of the parameters, the transition is first order at those values. 
If it grows sublinearly as $\sim N^\eta$, with a power $0<\eta<1$, or logarithmically, the transition is of the second order. 

In practice, it is also convenient to use the Pearson correlation coefficient \cite{Pearson_1896}
\begin{equation}
\label{eq:rho}
\rho_{pm} =
\frac{\chi_{pm}}{\sqrt{\chi_p \chi_m}}
\end{equation}
to examine the correlations between $P$ and $M$. By construction, $\rho_{pm}$ takes values in the range $[-1,1]$. 

\section{Methods}

We will use three different methods to investigate the behavior of the model: Monte Carlo simulations (Sect.~\ref{sec:MC}); the exact solution for small systems based on exact enumeration of the partition function and statistical averages (Sect.~\ref{sec:SS}); the analytic solution of limiting cases (Sect.~\ref{sec:ALC}). 

\subsection{Monte Carlo simulations \label{sec:MC}}

We use a Markov chain Monte Carlo method based on heat-bath updates. A single transition $t\rightarrow t+1$ in the Markov chain relies on randomly selecting an edge $ab$ and choosing its new sign: $s_{ab}(t+1)=s=\pm 1$ according to the probability distribution
\begin{equation} \label{hb_general}
\Prob(s_{ab}(t+1)=s) = \frac{\exp\left(-\beta U(s,\{s(t)\}_r)\right)}{\sum_{s=\pm 1} \exp\left(-\beta U(s,\{s(t)\}_r)\right)},
\end{equation}
where $U(s,\{s\}_r)$ is the energy of the system: the first argument of $U$ is the sign of edge $ab$, and the second argument, $\{s\}_r$, 
is the collection of signs of all the remaining edges, which do not change when the sign of $ab$ is updated.
This is a general equation, but it can be simplified because most of the factors in the numerator and denominator cancel out.
What remains is the dependence on the nearest neighborhood of the updated edge. 
The edge $ab$ belongs to two elementary loops.
For a square lattice, each loop consists of four edges; for example, $\{ab,bc,cd,da\}$ and $\{ab, be,ef,fa\}$. 
Eq.~\eqref{hb_general} simplifies to 
\begin{equation}
\label{eq:evol_p}
\Prob(s_{ab}(t+1)=s)=\frac{
\exp\left(s \xi_{ab}(t)\right)}
{\sum_{s=\pm 1} \exp\left(s \xi_{ab}(t)\right)},
\end{equation}
where 
\begin{equation} \label{staple4}
    \xi_{ab} = \varepsilon'\left(s_{bc} s_{cd} s_{da} + s_{be}s_{ef}s_{fa}\right)+ h'.
\end{equation}
We see again that the model depends on $\beta$, $h$ and $\varepsilon$ via effective parameters: $h'=\beta h$ and $\varepsilon'=\beta \varepsilon$.
For a triangular lattice, $\xi_{ab}$ \eqref{staple4} 
has to be replaced with 
\begin{equation} \label{staple3}
    \xi_{ab} = \varepsilon'\left(s_{bc} s_{ca} + s_{bd}s_{da}\right)+h',
\end{equation}
where $ab$ is a common edge of $\{ab,bc,ca\}$ and $\{ab,bd,da\}$~\cite{1911.13048}.
The term $\xi_{ab}$ can be identified as a local field acting on $s_{ab}$ \cite{1911.13048,2206.14226,2407.02603}.

A natural unit of the Monte Carlo time is a sweep, which corresponds to $N_e$ edge updates. Monte Carlo runs are organized as follows.  We first perform thermalization sweeps, starting from a randomly selected initial state. After the thermalization phase, we continue with the simulations, measuring the observables every sweep. The number of thermalization sweeps is on the order of $10^5$, and so is the number of sweeps with measurements. We perform simultaneously $64$ independent simulations, so there are $64\times 10^5$ measurements.

We simulate systems on a toroidal lattice of dimensions $L\times L$ with periodic boundary conditions. In most simulations $L=16$ and $N=L\times L =256$. 

For large $\varepsilon'$, the algorithm has a relatively low acceptance rate, which means that very often the states $s_{ab}$ remain unchanged by updates. In effect, the consecutive configurations generated by the algorithm are highly correlated, and the Markov chain explores the state space very slowly. To speed up this process, one can supplement the standard scheme based on edge updates with the following node updates. In a single node update, a node $a$ is selected at random, and then new values $s_{ab}$ are assigned to all edges $ab$ emerging from $a$
\begin{equation} \label{sch}
 s_{ab}(t+1) = \sigma_a s_{ab}(t),
\end{equation}
where $\sigma_a=\pm 1$ is an auxiliary random variable
associated with the node $a$. This variable is selected with the probability
\begin{equation} \label{hbsigma}
\Prob(\sigma_a=\sigma) = \frac{\exp\left(\sigma \zeta_{a}\right)}{
\sum_{\sigma=\pm 1} \exp\left(\sigma \zeta_a\right)},
\end{equation}
where 
\begin{equation}
\zeta_a = h' \sum_{\{ab\}} s_{ab}.
\end{equation}
The sum is over all edges $\{ab\}$ emerging from $a$.
In the case of a square lattice there are four such edges, and in the case of a triangular lattice there are six.
It is easy to see that changing the sign of edges \eqref{sch} simultaneously does not change the parity of any loop because an even number of edges changes the sign in each loop. Therefore, the term $P$ \eqref{P} in the energy is independent of $\sigma_a$ \eqref{sch}.
The only part of the energy \eqref{U} that changes is that related to $M$ \eqref{M}, and this part shapes the heat-bath probability \eqref{hbsigma}. 
For $h=0$, the updates \eqref{sch} correspond to microcanonical
changes that preserve the energy of the system.
In the limit $\varepsilon'\to \infty$, this algorithm becomes exact because, in this limit, $p=1$ \eqref{eq:p} is constant. This limit corresponds to structural balance. It is not a coincidence that Eq.~\eqref{hbsigma} resembles a heat bath equation for the nearest-neighbor Ising model. In the limit $\varepsilon'\to \infty$, the model is equivalent to the nearest neighbor Ising model \cite{2512.02644}.

\subsection{Exact enumeration for small systems \label{sec:SS}}

The statistical average of any observable 
$A=A(\{s_{ab}\})$ 
can be calculated directly from the formula
\begin{equation}
\langle A \rangle = \frac{1}{Z} \sum_{\{s_{ab}\}} A 
\exp\left(\varepsilon' P + h'M\right),
\end{equation}
where $Z$, $P$, and $M$ are given by \eqref{Z}, 
\eqref{P}, and \eqref{M}, respectively. Choosing $A=P,M,P^2,M^2,PM, \ldots$, we can, in particular, calculate the average densities \eqref{eq:pm}, the susceptibilities \eqref{eq:chis}, and {\em etc.} The method is straightforward---the only limitation is the number of terms in the sum, which is $2^{N_e}$. We will compute the average densities and susceptibilities for different values of the parameters $h'=\beta h$ and $\varepsilon'=\beta\varepsilon$ for a $3\times3$ square lattice, for which there are $2^{18}$ states (accounting for periodic boundary conditions adds 6 boundary edges to the 12 internal edges that connect the 9 lattice nodes, bringing the total number of edges to $N_e=18$), and for a $2\times2$ triangular lattice, for which there are $2^{12}$ states (imposing periodic boundary conditions supplements the 5 internal edges with 7 boundary edges, resulting in $N_e=12$). In both cases, the calculations are feasible.

\subsection{Analytic calculations for limiting cases \label{sec:ALC}}

The free energy density \eqref{fe_density} can be found analytically in special cases, including: (1) 
$\varepsilon'=0$ (Sect.~\ref{sec:eps=0}); (2) $\varepsilon' \to \infty$ (Sect.~\ref{sec:epstoinfty}), and (3) $h'=0$ (Sect.~\ref{sec:h=0}). So let us consider these three cases in turn.

\subsubsection{Limiting case $\varepsilon'=0$ \label{sec:eps=0}}

For $\varepsilon'=0$, the system consists of non-interacting signs $s_{ab}$ of edges, so the partition function is 
\begin{equation}
    Z = \prod_{\{ab\}} \sum_{s_{ab}=\pm 1} e^{\beta h s_{ab}/n_e} = \left(2\cosh(\beta h/n_e)\right)^{N_e}.
\end{equation} 
The average magnetization \eqref{eq:m} is $m=\tanh(\beta h/n_e)$, and the susceptibility \eqref{eq:chim} $\chi_m = \beta/n_e \cosh^{-2}(\beta h/n_e)$. 
It follows from the independence of edge spins that $p = m^4$ and that \eqref{eq:chip}
\begin{equation} \label{chip4}
  \chi_p = \frac{\beta}{n_p} \left((1 - m^8) + 4m^6(1-m^2)\right).
\end{equation}
The first term in the parentheses comes from overlapping plaquettes, while the second comes from plaquettes sharing an edge. 
For a square lattice $n_p=1$, thus $n_p$ can be skipped in the last equation.
The corresponding expression for the triangular lattice is 
\begin{equation} \label{chip6}
  \chi_p = \frac{\beta}{n_p} \left((1 - m^6) + 3m^4(1-m^2)\right)
\end{equation}
with $n_p=2$.

\subsubsection{Limiting case $\varepsilon'\to\infty$} \label{sec:epstoinfty}

Now we consider the case of $\varepsilon'\rightarrow \infty$.
In this limit $p$ tends to $p=1$ and $\chi_p$ to $\chi_p=0$; all loops have positive parity $\pi(\gamma)=1$ \eqref{parity},
and the system is structurally balanced.
However, the edge spins are still free to take on values $\pm 1$, provided that the product of these values for any closed loop is equal to unity. The requirement that for any (elementary) loop the product is equal to unity can be viewed as a constraint imposed on edge spins. The main observation is that the constraint is automatically satisfied if, for any edge $ab$, $\sigma_{ab}$ is given by
\begin{equation} \label{sss}
 s_{ab} = \sigma_a \sigma_{b},
\end{equation}
where $\sigma_a,\sigma_b=\pm 1$ are spin variables associated with the nodes of the lattice, and $a$ and $b$ in Eq.~\eqref{sss} are endpoints of the edge $ab$ \cite{2512.02644}. It is obvious that the parity of any loop is one, $\pi(\gamma)=1$ \eqref{parity}, when the edge signs are constructed according to
the rule \eqref{sss}, since each $\sigma_a$ appears in the product an even number of times.

For $\varepsilon'\to\infty$, the partition function~\eqref{Z} reduces to the sum over node spins $\sigma_a$, which is equal to the partition function of the nearest neighbor Ising model without an external field:
\begin{equation} \label{equiv}
   Z\exp\left(-\beta \varepsilon N\right) \rightarrow Z_{\rm Ising} = \sum_{\{\sigma_a\}} \exp\left(\frac{\beta h}{n_e} \sum_{(ab)} \sigma_{a} \sigma_b \right).
\end{equation}
The normalization factor $\exp(-\beta \varepsilon N)$ cancels the trivial contribution from the term $P$ in $U$ \eqref{U}, which tends to $P=N$ in the limit $\varepsilon'\rightarrow \infty$.
For the purposes of the argument below, it is important that this normalization factor is independent of $h$.

The parameter $h$, which is an external field in the original model \eqref{Z}, plays the role of the spin coupling constant in the corresponding Ising model. The free energy density of the Heider model (in structural balance) is equal to the free energy density of the Ising model
\begin{equation} 
   \lim_{\varepsilon \rightarrow \infty} \left(f(\beta,\varepsilon,h) + \varepsilon \right) =  f_{\rm Ising}(\beta,h/n_e)
\end{equation}
up to an $h$-independent term $\varepsilon$. This term becomes infinite in the limit but can be subtracted before the limit is taken. 
This term gives zero when we calculate the derivatives of the free energy density with respect to $h$. The free energy density for the Ising model on a square lattice $f_{\rm Ising}(\beta,h)$ is known analytically \cite{Onsager_1944}. 
Using this analytic expression and setting $n_e=2$ (for the square lattice), we find \eqref{eq:m}  
\begin{equation} 
   m = \frac{1}{2} \coth(\beta h) \left(1 + \frac{2}{\pi} b K(a)\right),
   \label{eq:m-sq-epsinf}
\end{equation}
where $a= 2\sinh(\beta h)/\cosh^2(\beta h)$, $b= 2 \tanh^2(\beta h) - 1$, $a^2+b^2=1$; $K(a)$ is an elliptic integral of the first kind:
\begin{equation}
    K(a) = \int_0^{\pi/2} \frac{d\theta}{\sqrt{1-a^2\sin^2(\theta)}}.
\end{equation}
Taking one more derivative with respect to $h$, we find \eqref{eq:chim} 
\begin{equation} 
\begin{split}
   \chi_m = & \frac{\beta}{2\pi}\coth^2(\beta h)
 \left[2K(a) - 2E(a) - (1-b) \left(\frac{\pi}{2}+b K(a) \right)\right],
\end{split} \label{eq:chim_exact}
\end{equation}
where $E(a)$ is an elliptic integral of the second kind:
\begin{equation}
    E(a) = \int_0^{\pi/2} d\theta \sqrt{1-a^2\sin^2(\theta)}.
\end{equation}
A closer examination of the expression \eqref{eq:chim_exact} shows that $\chi_m$ is logarithmically divergent---{\em i.e.}, $\chi_m \sim \ln| \beta h - h'_{cr}|$---for $\beta h \rightarrow h'_{cr}\equiv  \ln (1 + \sqrt{2}) \approx 0.88137$. The divergence comes from the elliptic integral $K(a)$ for $a\rightarrow 1$ in Eq.~\eqref{eq:chim_exact}. The divergence indicates that there is a second-order phase transition at $h\beta=h'_{cr}$.

We emphasize that the equivalence of the Heider model for $\varepsilon'\rightarrow \infty$ with the Ising model maps the magnetic susceptibility of the Heider model to the heat capacity of the Ising model.
Thus, the logarithmic singularity of the magnetic susceptibility of the Heider model at the critical point is the mirror image of the logarithmic singularity of the heat capacity of the Ising model. 

\subsubsection{Limiting case $h=0$ \label{sec:h=0}}

A common feature of the two limits $\varepsilon\to 0,\infty$ is that, in both $\chi_m=\beta/n_e$, for $h=0$.
We will show below that this is a manifestation of the general feature of the model, namely that $\chi_m=\beta/n_e$ for $h=0$, regardless of the value of $\varepsilon$, being a consequence of the local symmetry that appears in the model for $h=0$. The expression for energy \eqref{U} reduces to an 
expression
\begin{equation}
   U = -\varepsilon P, 
     \label{Uh=0}
\end{equation}
which is invariant with respect to the local transformation
at a node $a$
\begin{equation} \label{gauge}
   s_{ab} \rightarrow -s_{ab} 
\end{equation}
applied simultaneously to all edges $ab$ sharing the node $a$. 
This is a local symmetry of the model because it holds for any node $a$. It is sometimes called gauge symmetry \cite{Kogut_1979}.
The transformation \eqref{gauge} can be applied to multiple nodes simultaneously without changing the energy \eqref{Uh=0}. It is equivalent to a microcanonical
update rule \eqref{sch} for $h=0$ that we mentioned while discussing the
Monte Carlo algorithm. 
From this invariance, it follows that $\langle s_{ab}\rangle = -\langle s_{ab}\rangle$, hence $\langle s_{ab} \rangle = 0$.
Similarly, we see that $\langle s_{ab} s_{cd}\rangle = 0$ for any two distinct edges $ab$ and $cd$ (including edges having a common node, {\em e.g.}, $\langle s_{ab} s_{bd}\rangle = 0$). 
This means that the edge sign values are uncorrelated. As a consequence, the magnetization $M = 1/n_e \sum_{(ab)} s_{ab}$ is a sum of $N_e$ uncorrelated values $\pm 1$, multiplied by $1/n_e$. Therefore $\langle M \rangle =0$, and $\langle M^2 \rangle = N_e/n_e^2=N/n_e$, and hence
$m=0$ \eqref{eq:m} and $\chi_m=\beta/n_e$ \eqref{eq:chim}, for any $\varepsilon$ provided $h=0$. 

\begin{figure}
\includegraphics{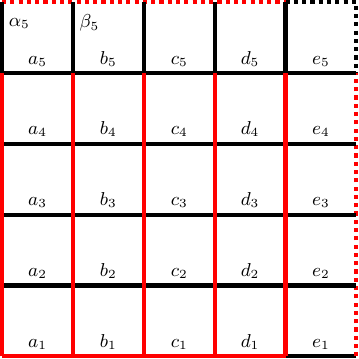}
\caption{\label{fig:sq_g} A $5 \times 5$ square lattice with periodic boundary conditions.
Periodic copies of edges are shown as dotted lines.
Spins of edges belonging to the spanning tree (in red) are fixed $s_{ab}=1$. 
Remaining spins (black edges) are effective degrees of freedom in Eq.~\eqref{Uh=0} under this gauge fixing. Due to the gauge fixing, interactions in each column take the form of 1D Ising interactions, for instance in the first column they are $s_{a_2} + s_{a_2} s_{a_3} + s_{a_3}s_{a_4} + s_{a_4}s_{a_5}$, in the fifth $s_{e_1} s_{e_2} + s_{e_2} s_{e_3} + s_{e_3}s_{e_4} + s_{e_4}s_{e_5}$.
The only exception is the top row, where interactions involve products of three or four spins per plaquette, for example the contribution of the top left plaquette is $s_{\alpha_5} s_{a_5} s_{\beta_5}$.}
\end{figure}

Another consequence of the invariance of $U$ \eqref{Uh=0} under the transformation \eqref{gauge} is that $u = \tanh(\beta \varepsilon/n_p)$ \eqref{eq:p} and $\chi_p = \beta/n_p \cosh^{-2}(\beta \varepsilon/n_p)$ \eqref{eq:chip}. To see this, we first note that we can set (gauge) the value $s_{ab}$ on any edge $ab$ to unity, using the transformation \eqref{gauge}, without changing the energy \eqref{Uh=0}. Then we can set the value of $s_{bc}$ to $s_{bc}=1$ of another link emerging from $b$, then we can do the same for a link emerging from $c$ {\em etc.}, and we can do this as long as the gauged links do not form a loop. In other words, we can set the values of all edge spins $s_{ab}$ in the spanning tree to be $s_{ab}=1$. For a square lattice with periodic boundary conditions,
the spanning tree can be chosen to be a comb graph, which consists of a horizontal shaft with $L$ edges and $L+1$ vertical teeth, each having $L-1$ edges, as in Fig.~\ref{fig:sq_g}. 
We see that all plaquettes between the teeth have only one or two free edge spins, which form 1D Ising interactions with the coupling constant $\beta/n_p$ (see Fig.~\ref{fig:sq_g}). The exceptions are plaquettes in the top row, but the number of these plaquettes is $L$, which is much smaller than the number of plaquettes between the teeth, $L(L-1)$. In effect, the contribution of the top row plaquettes to the energy density $u=U/L^2$ is of order $\mathcal{O}(1/L)$ and thus it vanishes in the limit $L\rightarrow \infty$. What remains is the contribution from 1D Ising interactions coming from plaquettes lying between the teeth of the spanning tree, which gives $u = \tanh(\beta \varepsilon/n_p)$ in the limit $L\rightarrow \infty$. The procedure of gauge fixing for edges on a spanning tree can be repeated for any two-dimensional lattice, in particular for a triangular lattice, where it also gives the same result $u = \tanh(\beta \varepsilon/n_p)$ in the thermodynamic limit.

\section{Results}

The results will be presented in the form of graphs showing how the average densities \eqref{eq:pm}, the susceptibilities \eqref{eq:chis}, and the correlation coefficient \eqref{eq:rho} depend on $\beta h$ for different values of the parameter $\beta \varepsilon$. 
In fact, we found it more convenient to use the rescaled parameters $\beta h/n_e$ and $\beta \varepsilon/n_p$ in the comparative study for triangular and square networks, because for these parameters, the range in which interesting features of the results can be observed is approximately the same in both cases.

The Monte Carlo results will be shown as dashed lines in different colors corresponding to different $\beta \varepsilon$, as indicated in the figure legends. 
The error bars are smaller than the line width, so they are not displayed. The results for small systems will be displayed as thin lines in the same colors as the corresponding Monte Carlo results.

The analytic results for $\varepsilon=0$, Sect.~\ref{sec:ALC}, will be displayed as thick, transparent yellow lines. The yellow line is difficult to see in the graphs because it perfectly matches the numerical results, which are drawn in black for $\varepsilon=0$.

\begin{figure}
\includegraphics[width=\columnwidth]{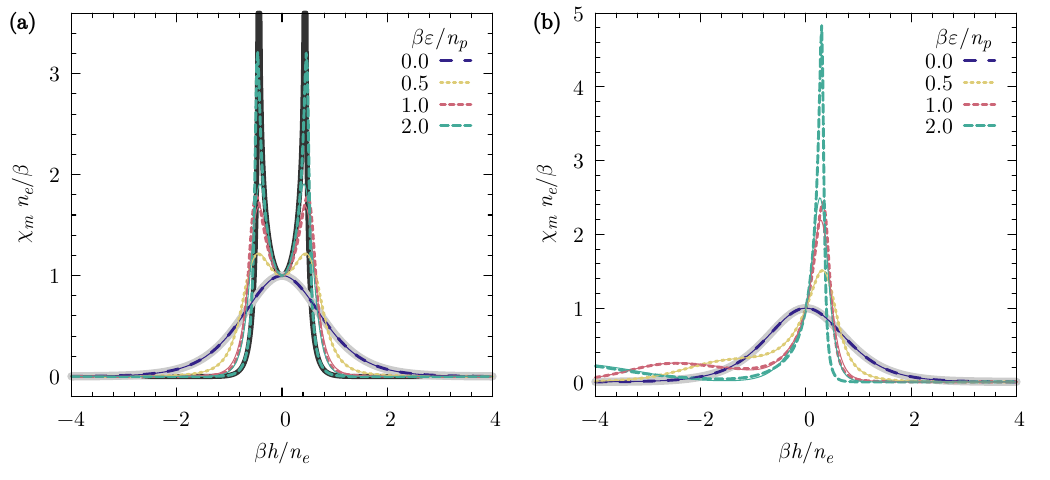}
\caption{\label{fig:chi_m}
Susceptibility $\chi_m n_e / \beta$ \eqref{eq:chim} for (a) square lattice, (b) triangular lattice. 
$\chi_m$ is a measure how sensitive the mean relation is with respect to a change of the social field $h$. 
Thick light-gray lines represent analytic solution for $\varepsilon=0$, $\chi_m n_e / \beta = \cosh^{-2}(\beta h/n_e)$.
Thick black in (a) is used for the analytic solution in the limit of $\varepsilon \to \infty$ \eqref{eq:chim_exact}.
Dashed lines are used for Monte Carlo results for $16 \times 16$ lattices, and thin lines in the corresponding colors for the small-systems ($N_p=9$ squares; $N_p=8$ triangles) exact results.
All curves take the value one at the zero point. 
For $\varepsilon\to \infty$ and $N\to \infty$, the curves diverge at the critical points $\beta h/n_e = \pm \ln (\sqrt{2}+1)/2\approx \pm 0.44069$ for the square lattice and $\beta h/n_e = \ln (3)/4 \approx 0.27465$ for the triangular lattice.}
\end{figure}

Additionally, for the square lattice, the results for $m$ and $\chi_m$ in the limit $\varepsilon\to\infty$ will be displayed as transparent light-green lines. They arise from the equivalence with the Ising model and the Onsager solution for the latter, Sect.~\ref{sec:ALC}.
For the triangular lattice, the solution is not known, but the critical value is known $\beta h/n_e = \ln(3)/4\approx 0.27465$ \cite{Baxter_1982}. The position of the susceptibility maximum for large $\varepsilon$ is consistent with this value (see Fig. \ref{fig:chi_m}). 

\begin{figure}
\includegraphics[width=\columnwidth]{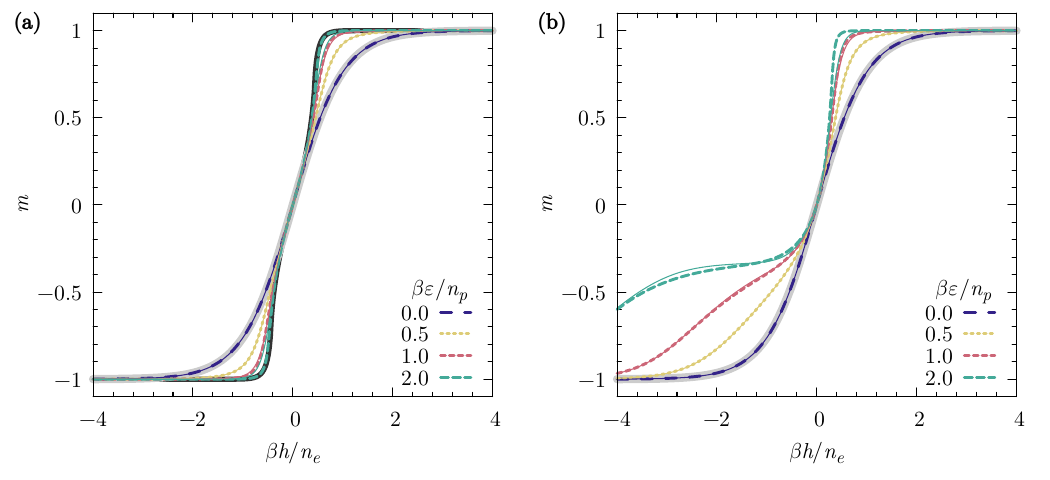}
\caption{\label{fig:m} Average density $m$ \eqref{eq:m} vs. $\beta h/n_e$
for different values of the parameter $\beta \varepsilon/n_p$ for (a) square lattice, (b) triangular lattice. 
The social interpretation of $m$ is the mean value of the interpersonal relations.
Monte Carlo results for $16 \times 16$ lattices are shown as dashed lines in colors corresponding to different values of $\beta \varepsilon/n_p$. 
The results for small systems  ($N_p=9$ squares; $N_p=8$ triangles) are displayed as thin lines in the same colors as the corresponding Monte Carlo results. 
Thick light-gray line represents the analytic solution $m=\tanh(\beta h/n_e)$ for $\varepsilon=0$ and thick black in (a) is used for the analytic solution for $\varepsilon\to\infty$ \eqref{eq:m-sq-epsinf}. 
The tangent of all curves at the zero point is one. For $\varepsilon \to \infty$ and $N\rightarrow \infty$, the tangent at the critical points $\beta h/n_e = \pm \ln (\sqrt{2}+1)/2 \approx \pm 0.44069$ for the square lattice and $\beta h/n_e = \ln (3)/4 \approx 0.27465$ for the triangular lattice is infinite.
}
\end{figure}

Fig.~\ref{fig:m} compares the average density $m$ \eqref{eq:m} for (a) the square lattice and (b) the triangular lattice. 
The main difference between the two cases is the symmetry: for the square lattice, $m(h)$ is an odd function, $m(-h)=-m(h)$; while for the triangular lattice, it is an odd function only for $\varepsilon=0$. This is related to the symmetry of $U$ \eqref{U} and $M$ \eqref{M} under the simultaneous transformation of all signs $s_{ab}\rightarrow \tilde{s}_{ab}=-s_{ab}$ and the field $h\rightarrow \tilde{h}=-h$. 
For the square lattice $U \rightarrow \tilde{U}=U$ and $M\rightarrow \tilde{M}=-M$. For the triangular lattice $U \rightarrow \tilde{U} \ne U$ and $M\rightarrow \tilde{M}=-M$, except for $\varepsilon=0$, for which $U \rightarrow \tilde{U}=U$ and $M\rightarrow \tilde{M}=-M$ also apply. Thus, for the square lattice $\langle M(-h) \rangle = -\langle M(h)\rangle$ and the triangular lattice $\langle M(-h) \rangle \ne -\langle M(h)\rangle$, except for $\varepsilon=0$, for which $\langle M(-h) \rangle = -\langle M(h)\rangle$.

A common feature of all the curves $m$ vs. $\beta h/n_e$ is that the tangent at the zero point is equal to one. This follows from the local symmetry at $h=0$,
as discussed in Sect.~\ref{sec:ALC}. 

We note that the results for small systems (thin lines), as discussed in Sect.~\ref{sec:SS} do not differ significantly from the Monte Carlo results obtained for networks of size $16\times 16$, Sect.~\ref{sec:MC} (dashed lines), showing that the effect of finite size is rather small. In fact, only for $\beta \varepsilon/n_p =2$ do we see with the naked eye that the thin red line deviates from the dashed red line. 

We also note that the results for $\varepsilon=0$ are perfectly consistent with the theoretical prediction (yellow transparent line). This will be the case for all other physical quantities discussed below.

The analytic solution for an infinitely large square lattice discussed in Sect. \ref{sec:ALC} (transparent green line) is very close to the results for $\beta \varepsilon/n_p =2$ for the dimensions $16\times 16$, which means that the system is already very close to the structural balance for $\beta \varepsilon =2$. 

Finally, we note that the curves in Fig.~\ref{fig:m} become steeper near the phase transition points as the parameter $\beta \varepsilon/n_p$ increases. For $\varepsilon\rightarrow \infty$, in the thermodynamic limit $L\rightarrow \infty$, the tangent to the curve is exactly vertical at the critical point (green transparent line). The derivative at this point is infinite and corresponds to a logarithmic divergence of the susceptibility at the critical point.

\begin{figure}
\includegraphics[width=\columnwidth]{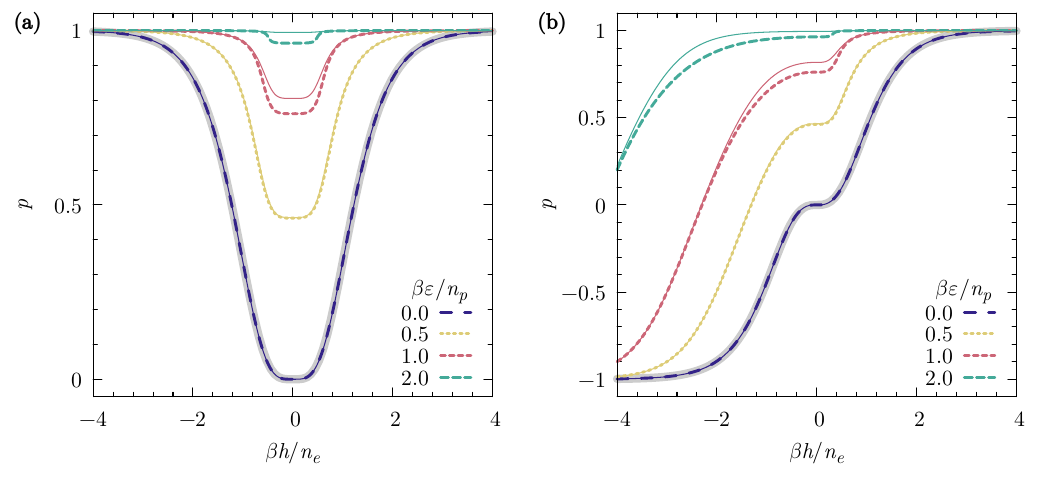}
\caption{\label{fig:p}
Average density $p$ \eqref{eq:p} for (a) square lattice, (b) triangular lattice. 
According to the sociological interpretation, $p$ is a mean contribution of  elementary closed loops to the balance.  
Dashed lines are used for Monte Carlo results for $16 \times 16$ lattices and different values of $\beta \varepsilon/n_p$; thin lines in the corresponding colors show the exact results for the small-systems ($N_p=9$ squares; $N_p=8$ triangles).
Thick light-gray lines represent analytic solutions for $\varepsilon=0$: (a) $p=\tanh^4(\beta h/n_e)$, and (b) $p=\tanh^3(\beta h/n_e)$.
}
\end{figure}

Fig.~\ref{fig:p} compares the average density $p$ \eqref{eq:p} for square and triangular lattices. Repeating the arguments about the symmetry under the simultaneous transformation of all signs $s_{ab}\rightarrow \tilde{s}_{ab}=-s_{ab}$ and the field $h\to \tilde{h}=-h$, we immediately find that (a) for the square lattice, $p(h)$ is an even function: $p(-h)=p(h)$, and (b) for the triangular lattice, $p(h)$ is an odd function: $p(-h)=-p(h)$ for $\varepsilon=0$, and otherwise is neither an odd nor an even function. 

For $\varepsilon=0$, the results are perfectly consistent with the theoretical predictions. There is some finite-size dependence for higher values of $\beta \varepsilon/n_p$. In the case of a square lattice, for $\beta \varepsilon/n_p = 2$, the average density $p$ for the system $16\times 16$ drops below unity for small $\beta h$, indicating that the system is not fully structurally balanced.
In other words, there is a non-vanishing probability of finding an unbalanced elementary loop $\gamma_e$ having a negative parity: $\pi(\gamma_e)=-1$. The probability drops to zero when $\beta \varepsilon$ increases.

\begin{figure}
\includegraphics[width=\columnwidth]{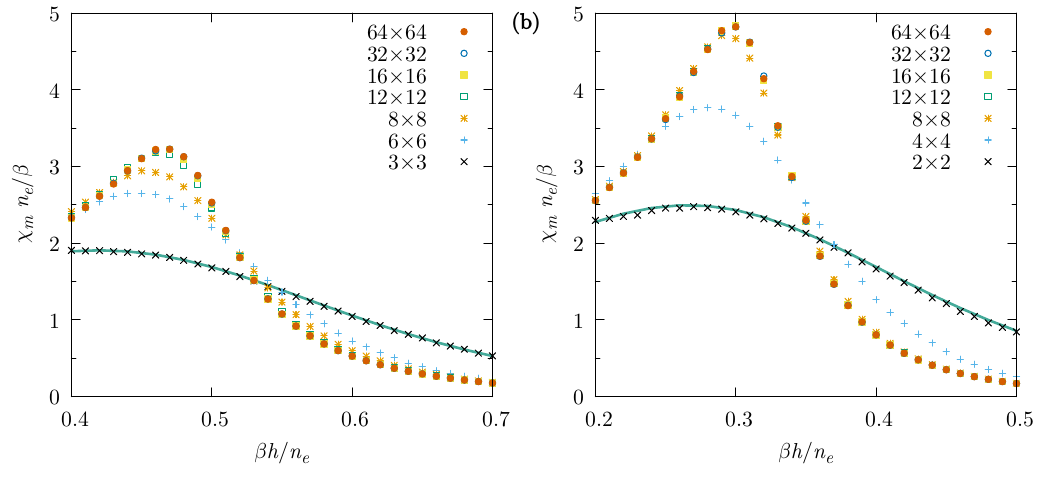}
\caption{\label{fig:chi_m_sizes}
Susceptibility $\chi_m n_e / \beta$ for $\beta \varepsilon / n_p = 2$ and various sizes $L \times L$ of (a) square lattice, (b) triangular lattice.
Discrete points are used for the Monte Carlo results, and the green line indicates the exact results obtained for the small-systems ($N_p=9$ squares in $3 \times 3$ lattice; $N_p=8$ triangles in $2 \times 2$ lattice).}
\end{figure}

Fig.~\ref{fig:chi_m} shows the edge spin susceptibility $\chi_m$ \eqref{eq:chim}.  For a square lattice, it is an even function of the field $h$, regardless of $\varepsilon$, while for a triangular lattice, it is an even function of the field $h$ only for $\varepsilon=0$. The curves have local maxima near the critical values of $\beta h/n_e =\pm \ln(\sqrt{2} +1)/2 \approx \pm 0.44069$ for the square lattice \cite{Onsager_1944, Codello_2010}, and $\beta h/n_e= \ln(3)/4 \approx 0.27465$ for the triangular lattice \cite{Baxter_1982, Codello_2010}. 
The maxima correspond to the largest fluctuations of edge spins.
In fact, they become infinite in the thermodynamic limit, where the susceptibility diverges logarithmically as 
$\beta h/n_e$ tends to the critical value. The transparent green line, which represents the exact result \eqref{eq:chim_exact}, extends to infinity at the critical point. In the case of a finite lattice, the maximum has a finite height, but this height increases logarithmically as the size of the system increases \cite{PhysRevLett.28.1516}.
We note that for $16\times 16$ lattice, the susceptibility already has a pronounced maximum, which follows the critical analytic solution for $\varepsilon\to\infty$ in the thermodynamic limit over a wide range; see Sect.~\ref{sec:ALC}. 

For infinite $\varepsilon$, the heights of the susceptibility maxima $\chi_m$ increase logarithmically with $L$, which follows from the equivalence with the heat capacity of the Ising model \cite{2512.02644}. However, for an arbitrarily large but finite $\varepsilon$, the height of the maxima approaches a constant independent of $L$ in the limit $L\rightarrow \infty$. In this limit, the maxima reach a limiting shape independent of $L$. 
This is illustrated in Fig. \ref{fig:chi_m_sizes},
where it is seen that the peak shape stabilizes with increasing $L$. The shape depends on $\varepsilon$; the larger $\varepsilon$, the higher the maximum. In summary, the system
is critical only for $\varepsilon=\infty$, at the critical value $h$.
For finite $\varepsilon$, no phase transition occurs.

\begin{figure}
\includegraphics[width=\columnwidth]{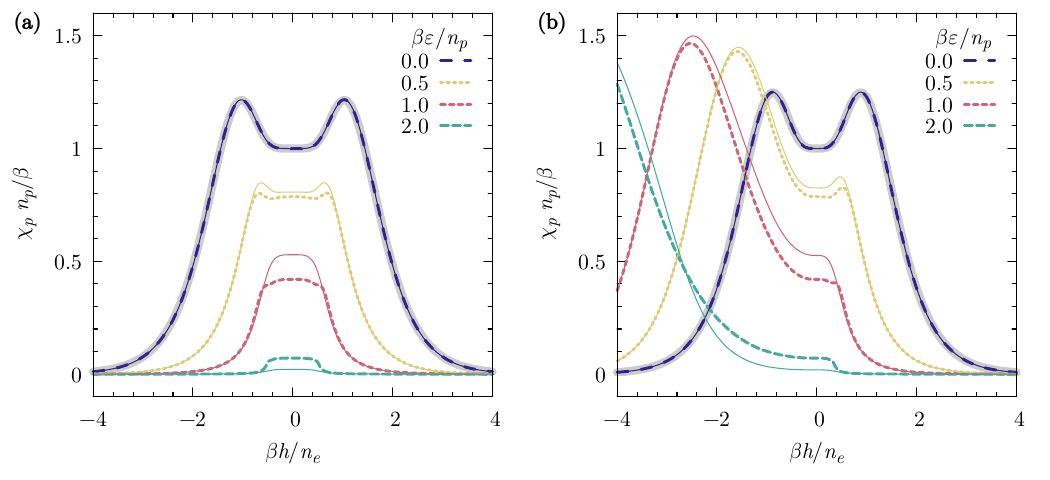}
\caption{\label{fig:chi_p}
Susceptibility $\chi_p n_p/ \beta$ \eqref{eq:chim} for (a) square lattice, (b) triangular lattice. 
$\chi_p$ is a measure how sensitive the balance is with respect to a change of the social field $h$. 
Thick light-gray lines represent analytic solutions for $\varepsilon=0$ given by Eqs.~\eqref{chip4} and \eqref{chip6}, dashed lines are used for Monte Carlo results for $16 \times 16$ lattices, and thin lines in the corresponding colors for exact results obtained  for the small-systems ($N_p=9$ squares; $N_p=8$ triangles).}
\end{figure}

Fig.~\ref{fig:chi_p} shows $\chi_p n_p/\beta$ vs. $\beta h/n_e$. For $\varepsilon\to 0$, the results are consistent with the analytical formulas given in Eqs.~\eqref{chip4} and~\eqref{chip6}. For the square lattice, $\chi_p n_p/\beta$, which is proportional to the variance of $p$, decreases as $\beta \varepsilon/n_p$ increases. 
In the limit $\varepsilon \rightarrow \infty$, the variance of $p$ drops to zero, and $p$ becomes equal to unity, which means that the system reaches a state of structural equilibrium. For large but finite $\varepsilon$, for example for $\beta \varepsilon/n_p =2$, there are still some non-zero fluctuations of $p$, seen in the plot as non-zero values of $\chi_p n_p/\beta$ for small values of the argument $\beta h/n_e$.

For the triangular lattice, the function $\chi_p$ is an even function of $h$ only for $\varepsilon =0$. When $\varepsilon$ increases, the maxima become asymmetric, and the graph shifts to the left. One can argue that the value of the susceptibility $\chi_p$ at any finite values of $h$ and $\beta$ will tend to zero in the limit $\varepsilon \to \infty$. This means that structural balance will be reached for any finite $h$ in the limit $\varepsilon\to \infty$. 
We note that the results for small systems ($3\times 3$ square lattice and $2\times 2$ triangular lattice) are fairly close to those for $16\times 16$. 

\begin{figure}
\includegraphics[width=\columnwidth]{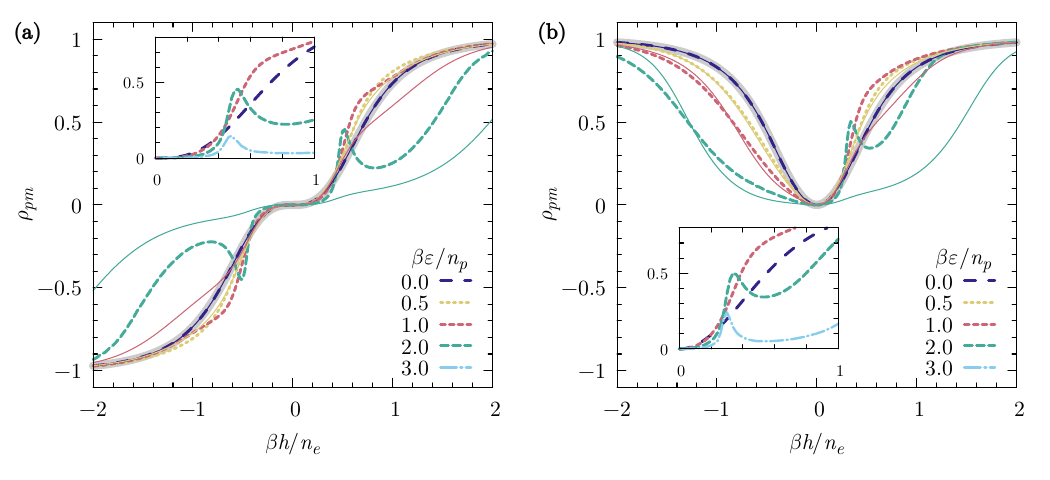}
\caption{\label{fig:rho_pm}
Correlation coefficient $\rho_{pm}$ \eqref{eq:rho} for (a) square lattice, (b) triangular lattice.
For the square lattice, strong negative field $h$ induces a positive contribution to the balance $p$, hence the correlation is negative there, in contrast to the triangular lattice.
Thick light-gray line represents analytic solution for $\varepsilon=0$, dashed lines are used for Monte Carlo results ($16 \times 16$ lattices), and thin lines for the small-systems ($N_p=9$ squares; $N_p=8$ triangles) exact results.
Insets illustrate the disappearing peak with an additional dash-dotted blue line for $\beta \varepsilon / n_p = 3$.}
\end{figure}

Fig.~\ref{fig:rho_pm} shows the Pearson correlation coefficient $\rho_{pm}$ \eqref{eq:rho}. Again, one can argue, using symmetry arguments, that $\rho_{pm}$ as a function of $\beta h$ is an odd function for the square lattice, and it is an even function only for $\varepsilon=0$ for the triangular lattice. For the square lattice, $m$ and $p$ \eqref{eq:pm} are anti-correlated for $h<0$ and positively correlated for $h>0$, and there is a characteristic plateau close to $h=0$, where they are uncorrelated or weakly correlated.
For the triangular lattice, $m$ and $p$ are positively correlated, except for $h=0$, where they are uncorrelated.

\begin{figure}
\includegraphics[width=\columnwidth]{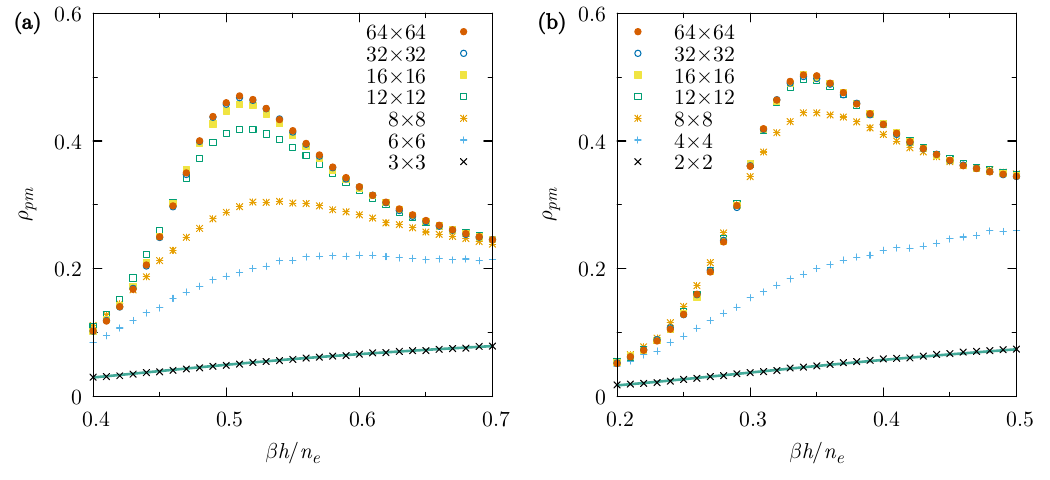}
\caption{\label{fig:rho_pm_sizes}
Correlation coefficient $\rho_{pm}$ for $\beta \varepsilon / n_p = 2$ and various sizes $L \times L$ of (a) square lattice, (b) triangular lattice.
Discrete points are used for the Monte Carlo results, and the green line indicates the exact results obtained for the small-systems ($N_p=9$ squares in $3 \times 3$ lattice; $N_p=8$ triangles in $2 \times 2$ lattice).}
\end{figure}

We note that while the correlation generally decreases with the increasing $\varepsilon$ (see insets in Fig.~\ref{fig:rho_pm}), this process is slower around the critical values of $\beta h / n_e$. 
It results in a characteristic peak for $\beta h/n_e$ close to the critical value, where the curve has a local extremum.
The peak is visible for larger lattices (Fig.~\ref{fig:rho_pm}). 
Its shape and height converge to a limit shape in the thermodynamic limit $L\rightarrow \infty$. 
This is illustrated in Fig.~\ref{fig:rho_pm_sizes}, which shows that the difference between the peak shapes decreases with the lattice size.
For $32\times 32$ and $64 \times 64$ lattices the shapes are practically identical.

\section{Discussion}

We performed a comparative analysis of the Heider model in
two dimensions on triangular and square lattices in the presence of an external field. The main difference between these two cases is that triads, which are elementary loops in the triangular lattice, change signs when all the signs of the edges on the lattice change simultaneously, and quadruples, which are elementary loops in the square lattice, do not change. From the point of view of the sign symmetry, these are the most basic examples of networks with only odd or only even elementary loops. 

In this study, we used three different computational methods: Monte Carlo simulations, exact calculations of partition functions and statistical averages for small systems by summing over all states, and analytical calculations of limiting cases. Using these methods, we obtained results that quantitatively illustrate the model behavior in different regimes. 
The methods, conceptually based on statistical mechanics, include random (`thermal') fluctuations derived from extremizing the Gibbs entropy \cite[p. 355]{Reichl_2016}. In our approach, the external field $h$ is an independent variable. A  step forward could be a construction of the mean-field approach, where the effective social field 
$h_\text{eff}$ contains also a contribution from the modified dependent variable $m$. Such a coupling should allow to capture hysteresis effects, as a history-dependent shift of the maximum of the susceptibility $\chi_m$ along the $h$-axis. However the effect is out of scope of this paper.

In this paper, by direct comparison of the results for $\varepsilon \to \infty$ with Onsager's analytical solution for the Ising model on a square lattice, we provided further evidence for the recent finding \cite{2512.02644} that the Heider model is equivalent in the limit $\varepsilon\rightarrow \infty$ to the Ising model with nearest-neighbor interactions.
A consequence of this equivalence is that the Heider model in two dimensions has a second-order phase transition in the field $h$, for $\varepsilon \to \infty$, where the susceptibility associated with the response to this field diverges logarithmically. 

With respect to signed social networks, the results allow us to assess the average sign and value of interpersonal relationships $m$, the balanced nature of the network $p$, and the sensitivity to the social field and stochastic fluctuations generated by thermal noise. These sensitivities are encoded in the susceptibilities $\chi_m$ and $\chi_p$. In particular, peaks of susceptibility appear in states in which two competing mechanisms---conformity to the social field, which may represent a social norm \cite{link1,link2}, and the tendency to eliminate cognitive dissonance \cite{Heider}---are characterized by comparable intensity.

With the example of a social field as common trust \cite{link2}, mentioned above, we can expect that in Denmark the interpersonal relations 
are rather positive and only weakly depend on actual content of media. 
Similarly, in Peru the relations are expected to be negative and even less dependent on press campaigns. 
To find more fragility, we should investigate countries where the mean trust is intermediate, as in Canada or United Kingdom \cite{link2}. 
In terms of our model, the susceptibility $\chi_m$ should be larger there.

From the perspective of the Heider balance, the square lattice is a paradoxical structure, as the balance is possible there without any friendly relation. Still, in accordance of the definition of balanced state \cite{Cartwright_1956} two mutually hostile groups can appear in two interpenetrating lattices. In this example the key element is the lack of contact between friends. A sociological counterpart could be a prison, where inmates are placed in single cells. Even if mentally supported by prayers of their adherents, they contact only with hostile prison guards, who also do not interact with each other. A theoretical counterpart of this scenario can be carried out for a negative $h$, large enough. The system is balanced and therefore more stable. Consequently, we can expect that opinions of guards about prisoners and those of prisoners about guards are  fixed, strong and negative.
This extreme example underlines the role of topology of the social network. Actually, this argumentation applies also to other networks where the number of nodes in an elementary loop is even, as for example in a two-dimensional hexagonal lattice.

\begin{acknowledgments}
This research was supported by a subsidy from the Polish Ministry of Science and Higher Education.
\end{acknowledgments}

%
\end{document}